\documentclass[journal]{IEEEtran}
\IEEEoverridecommandlockouts
\usepackage{cite}
\usepackage{amsmath,amssymb,amsfonts}
\usepackage{algorithmic}
\usepackage{graphicx}
\usepackage{textcomp}
\usepackage{xcolor}
\usepackage{amsmath}
\usepackage{float}
\usepackage{gensymb}
\usepackage{caption}
\usepackage{subcaption}
\usepackage{hyperref}
\usepackage{placeins}

\usepackage{dblfloatfix} 

\def\BibTeX{{\rm B\kern-.05em{\sc i\kern-.025em b}\kern-.08em
    T\kern-.1667em\lower.7ex\hbox{E}\kern-.125emX}}
\begin{document}

\title{On the Dynamics and State Dependent Multiple Equilibria of a Post-Buckled Ultra-Flexible Inverted Pendulum on a Rotating Hub
\\
}

\author{\IEEEauthorblockN{Prasanna Gandhi*}
\IEEEauthorblockA{\textit{Professor, Mechanical Engineering} \\
\textit{Indian Institute of Technology, Bombay}\\
Mumbai, India \\
gandhi@me.iitb.ac.in}
\and
\IEEEauthorblockN{Dhruvi Joshi}
\IEEEauthorblockA{\textit
{Mechanical Engineering} \\
\textit{Indian Institute of Technology, Bombay}\\
Mumbai, India \\}
\and
\IEEEauthorblockN{ Vivek Natarajan}
\IEEEauthorblockA{\textit{ Professor, Systems and Controls} \\
\textit{Indian Institute of Technology, Bombay}\\
Mumbai, India \\}
\and
\IEEEauthorblockN{\phantom{Padding}}
\IEEEauthorblockA{\phantom{Padding line}\\\phantom{Padding line}\\\phantom{Padding line}}
\and
\IEEEauthorblockN{Ravit Anand}
\IEEEauthorblockA{\textit{Mechanical Engineering} \\
\textit{Indian Institute of Technology, Bombay}\\
Mumbai, India \\}
}

 \author{Prasanna Gandhi*,  Dhruvi Joshi, Vivek Natarajan, Ravit Anand
 \thanks{P. Gandhi is with the Faculty of the Department of Mechanical Engineering, Indian Institute of Technology, Bombay, India, gandhi@me.iitb.ac.in}
 \thanks{D. Joshi is with the Department of Mechanical Engineering, Indian Institute of Technology, Bombay, India}
 \thanks{V. Natarajan is with the Centre for Systems and Controls, Indian Institute of Technology, Bombay, India}
 \thanks{R. Anand is with the Department of Mechanical Engineering, Indian Institute of Technology, Bombay, India}
 }


\maketitle

\section*{Abstract}
\noindent Compliant element systems with ultra-large deformation display
rich nonlinear dynamics and pose challenging control
problems, which, when solved, could enable enhancements in several mechatronics applications, such as soft robotics, MEMS, and biomedical applications. This paper considers post-buckled dynamic analysis of an inverted ultra-flexible pendulum actuated by a rotary hub. We first derive a complete set of equations capturing the dynamics of the system, essential for control development, using the assumed modes method framework, considering ultra-large deformations. Constrained Lagrange formulation is used for the same. In the
perfect inverted configuration with zero hub angle, the buckled beam would display two symmetric stable equilibria and one unstable. However, as the hub angle changes on either side, the equilibrium positions shift, and eventually two of them
vanish, and we are left with only one stable equilibrium. We use the dynamic equations to characterize this interesting phenomenon, demonstrating the continuous state dependence of multiple equilibria. 
Furthermore, experimental counterparts of the equilibrium results are meticulously obtained and discussed. Moreover, simulation results capture the nonlinear dynamics of this system. Overall, the work establishes a solid mathematical foundation with a control-amenable model for futuristic ultra-compliant mechatronic systems.

\section{Introduction}
\noindent
Systems with flexibility or compliance are widely preferred due to several advantages, including lightweight, low power consumption, high speeds, natural energy storage, and a soft feel, among others \cite{10438059}. They find several applications in multiple domains spanning across aerospace \cite{https://doi.org/10.1002/aisy.202400785}, biomedical robotics \cite{10.1115/1.4049491, 4624584}, and MEMS\cite{8972908}. Applications in soft robotics \cite{https://doi.org/10.1002/admt.202100018, banerjee2018soft, Cianchetti2018} are becoming more and more popular in the research quest of biomimicking naturally efficient systems. However, the advantages mentioned above in these systems come at the cost of intrinsically complex nonlinear dynamics and associated control challenges \cite{10438059, 10438341, Benosman_LeVey_2004, Rigatos2018} especially when considering multi-body systems \cite{Rong2019}. Control and dynamics of single and multibody flexible systems considered in these survey papers are with non-buckled links with small deformations. 
\\
\\
Past literature has shown multiple instances of exploration regarding the inverted flexible pendulum on a cart system. Tang, Ren \cite{6075558} performed modeling and simulation of a flexible inverted pendulum without a tip mass on a cart using a floating frame of reference formulation and used a simple low-pass filter as a controller. A deeper analysis of the nonlinear nature of the system, including bifurcation and hysteresis, has been conducted, and various approaches for developing control strategies have been proposed by Santina in \cite{9303976}. Rapid stabilization methods were explored for a flexible inverted pendulum mounted on a rotational pin joint by Chu et al. in  \cite{CHU2023109895}. These systems did not consider any tip mass attached to the pendulum. 
\\
\\
Adding to the complexity of more realistic situations, researchers considered a flexible pendulum on a cart (normal and inverted) system with a tip mass \cite{10.1115/1.4026831}. The presence of gravity or axial loading leads to buckling in these systems, and this phenomena, coupled with large deformations, make modeling and analysis, especially for control purposes, further involved. Large deformations have been studied and modeled in the past using various methods, such as utilizing the Cosserat rod theory for developing an explicit non-linear spatial beam Adomian decomposition model \cite{10.1115/1.4067023} and utilization of elliptical integrals for large deflections \cite{10.1115/1.4023558}.
Emergence of multiple equilibria when the tip mass is gradually increased was demonstrated by Patil et al. in \cite{10.1115/1.4026831}, and an experimentally validated dynamic model considering large deformations was presented by Vyas et al. in \cite{10.1007/978-981-15-8049-9_62}. Modeling approaches, such as the elastica theory in the static domain and the assumed modes method (AMM) in the dynamic domain, were adopted for this purpose. The control of vibration in flexible beams and stabilization of unstable equilibria (post-buckled scenario) were the next challenges taken up by researchers. 
Control strategies developed in the past have evolved from vibration suppression using planned flywheel motion \cite{CHU2024117975} to approaches that include robust control frameworks \cite{FRANCO2018539} and energy-shaping methods \cite{GANDHI201627}.
\\

\noindent
Studies on post-buckled beams clarify how multistability arises from elastic energy wells. Santina et al.\cite{9303976} discussed the effect of stiffness k as a parameter for bifurcation, while \cite{10.1115/1.4026831} focuses on the existence of multiple equilibria based on the load ratio. Moreover, the rotating beam system has been modeled and analyzed using a simplified Euler-Bernoulli beam model, assuming small deflections using the Lagrange-AMM method \cite{10.1007/978-981-96-5527-4_57}. Studies have emphasized the dynamic stiffening, and geometric nonlinearities under rotation \cite{CHU2023109895}, developing a strategy to rapidly stabilize the vibrations of an inverted flexible pendulum on a hub, without a tip mass, by controlling the angle of the hub on which it is mounted in a manner that enhances effective damping\cite{CHU2023109895}. An adaptive boundary controller for a flexible beam on a hub with a tip mass in a gravity-free system has been developed for attitude tracking and vibration control \cite{Yao2025}. However, with a tip mass, especially one that exceeds the critical buckling load, the beam buckling presents multiple challenges in controlling the beam on the rotating hub. Although it could be useful in practical applications, to the best of the authors' knowledge, scientific exploration into post-buckled beam systems operated at the base by a rotating hub (essentially a buckled single-link manipulator) remains an open problem in the literature. 
\\
\\
This paper unfolds interesting fundamental characteristics of an inverted buckled flexible pendulum on a rotating hub system and proposes a control-amenable model. Compared to the cart-mounted translation case, hub rotation disturbs the alignment of the reference longitudinal axis with the gravity vector, breaking the symmetry. As a result, the system admits rotation state-dependent static equilibria, the number and stability of which vary with configuration parameters. Moreover, centrifugal and Coriolis contributions lead to added complexity in dynamic motion. 
\\
\\
\noindent 
To begin with, we establish a mathematical context for the problem and derive the equations governing the system's dynamics. Large deformations are primarily considered by imposing the constancy of the beam's length as a constraint within the Assumed Modes Method (AMM) framework. The following sections are dedicated to a detailed analytical, numerical, and experimental exploration of the evolution of multiple equilibria of a dynamic system with the hub angle as a bifurcation parameter. A critical angle beyond which two of the three equilibria vanish is characterized, along with its variation with the tip mass. Finally, a case of the dynamic simulation is presented. This analysis develops the necessary insights to be acquired before deploying buckled compliant links or mechanisms for applications, including soft robotics.  

\section{Dynamics of the system}
\label{chp:AMM} 
\noindent In this section, we derive the equations of motion for the system by approximating the beam as a finite degree of freedom system, but still considering large deformations. 
The derivation is inspired by that given in \cite{10.1115/1.4026831}, where the system is a buckled flexible pendulum with a tip mass mounted on a cart.
\subsection{System Definition}
\begin{figure}[h!]
    \centering
    \includegraphics[width=0.4\textwidth]{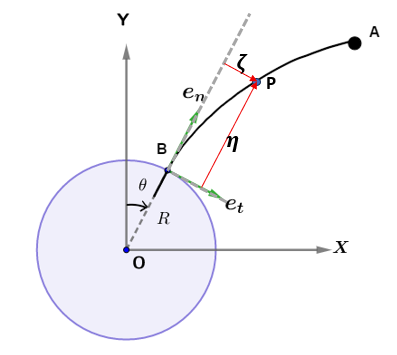}
    \caption{{Schematic of the rotary inverted ultra-flexible pendulum}}
    \label{fig:system}
\end{figure}
\noindent Fig.\ref{fig:system} describes our system where a beam BA with tip mass $M$ at A is fixed at B to a hub with radius R. Local coordinate system ${e_n}-{e_t}$, attached to hub at B as shown, is used to describe the beam. The geometrical relation between the local co-ordinates ${e_n}-{e_t}$  and the global frame $X-Y$ is given by:
\begin{subequations}
\begin{gather}
    \hat{e}_n(t)=\sin{\theta(t)}\hat{i}+\cos{\theta(t)}\hat{j}\\
    \hat{e}_t(t)=\cos{\theta(t)}\hat{i}-\sin{\theta(t)}\hat{j}
\end{gather}
\end{subequations}

\noindent The following notations for the generalized coordinates $\eta(l,t)$ (along $e_n$) and $\zeta(l,t)$ (along $e_t$) for any given point P on the beam (refer to Fig.\ref{fig:system}) will be used:
\begin{subequations}
\begin{equation}
    \dot{\eta}=\frac{\partial }{\partial t}\eta(l,t) \textbf{ ; } {\eta}'=\frac{\partial }{\partial l}\eta(l,t) \textbf{ ; } {\eta}''=\frac{\partial^2 }{\partial l^2}\eta(l,t) \textbf{ ; } 
\end{equation}
\begin{equation}
    \dot{\zeta}=\frac{\partial }{\partial t}\zeta(l,t) \textbf{ ; } {\zeta}'=\frac{\partial }{\partial l}\zeta(l,t) \textbf{ ; } {\zeta}''=\frac{\partial^2 }{\partial l^2}\zeta(l,t).
\end{equation}
\end{subequations}

\subsection{System Energy}
\noindent The total kinetic energy of the system can be given by 
\begin{equation}
    T=T_A+T_h+T_p
\end{equation}
where, the translational kinetic energy of the tip mass A is
\begin{equation}
    T_A=\frac{1}{2}M\left\Vert{\vec{{v}}_A}\right\Vert^2
    \end{equation} 
    and $\vec{v}_A$ = velocity of the tip .
\\The kinetic energy of the hub is given by
\begin{equation}T_h=\frac{1}{2}\mathcal{J}\dot{\theta}^2
\end{equation}
and the kinetic energy of the beam is expressed as
\begin{align}
T_p &= \int_0^L \frac{1}{2}\rho\Big[
    (\dot{\eta} - \zeta\dot{\theta})^2
     + \big((R + \eta)\dot{\theta} + \dot{\zeta}\big)^2
    \Big]\,\text{d}l.
\end{align}


\noindent The total potential energy of the system is given by
\begin{equation}
    U=U_A+U_p^e+U^g_p
\end{equation} where, the potential energy of the tip mass A is
\begin{equation}
    U_A=Mg((R+\eta(L,t))\cos{\theta}(t)-\zeta(L,t)\sin{\theta(t)})
\end{equation}
the elastic potential energy of the beam is given by
\begin{equation}
    \begin{split}
    U^e_p&=\frac{EI}{2}{\int_0^{L}}\frac{(\zeta''\eta'-\eta''\zeta')^2}{((\eta')^2+(\zeta')^2)^3}\text{d}l\\
    \end{split}
\end{equation}
and the gravitational potential energy of the beam is given by
\begin{equation}
    U^g_p=\int_0^L \rho g((R+\eta)\cos{\theta}-\zeta\sin{\theta})\text{d}l
\end{equation}
where, $\rho$ is the material density, $E$, the Young's modulus and $I$, the second moment of area.

\subsection{Assumed modes method (AMM)}
\noindent We apply Hamilton's principle to arrive at the equations of motion by approximating the distributed system with a finite degree of freedom system. The AMM essentially assumes total deformation as the sum of a few scaled mode shapes of the beam $\phi_i$ ($\eta$), where the scaling factor for each is the generalized coordinate $q_i(t)$. Thus, for the present case $\zeta $ can be expressed as a function of $\eta$ and t,
\begin{equation}
    \zeta(\eta,t)=\sum_{i=1}^n\phi_i(\eta)q_i(t)
\end{equation}
\noindent The modal function $\phi_i$ is obtained by solving the beam vibration equation (Euler-Bernoulli beam theory)\cite{meirovitch1986} using the variable separation method with the appropriate boundary conditions admissible in our case. The modal function is thus given by
\begin{equation}
\begin{split}
    {\phi}({\eta})=&L(\cosh{(\sigma_1{\eta}/L)}-\cos{(\sigma_2{\eta}/L)})\\&-L\bigg(\frac{\sigma_1^2\cosh{\sigma_1}+\sigma_2^2\cos{\sigma_2}}{\sigma_1^2\cosh{\sigma_1}+{(\beta L)}^2\sin{\sigma_2}}\bigg)\\&(\sinh{(\sigma_1{\eta}/L)}-\frac{\sigma_1}{\sigma_2}\sin{(\sigma_2{\eta}/L)})
\end{split}
\label{eq:modeshape}
\end{equation}
 where
\begin{equation}
    \sigma_1= \sqrt{-\frac{\bar{k}^2}{2}+\sqrt{\frac{\bar{k}^4}{4}+{( \beta L)}^4}} \text{ , } \sigma_2=\sqrt{\frac{\bar{k}^2}{2}+\sqrt{\frac{\bar{k}^4}{4}+( \beta L)^4}}  
\end{equation}
where,
\begin{equation}
\bar{k} = \sqrt{\frac{PE}{I}}L \text{ ; } P=\text{axial force acting on the beam} 
\end{equation} and
$\beta^4$ is the positive constant used to variably separate the equations
\begin{equation}\frac{1}{\phi(\eta)}\Big[\frac{d^4 }{d x^4}\phi(\eta)+\frac{P}{EI}\frac{d^2 }{d x^2}\phi(\eta)\Big]=\beta^4\end{equation}
\noindent Since our goal is to get equations of motion amenable to control development and still capture reasonably essential dynamics, we consider only the first mode in the analysis, i.e.
\begin{equation}
    \zeta(\eta,t)=\phi(\eta)q(t)
    \label{eq:firstmode}
\end{equation}
For the constrained Lagrangian formulation, the Lagrange function augmented with constraints is given by, 
\begin{equation}
    {L}_c:={L}(t,u_1,u_2,u_3,\dot{u}_1,\dot{u}_2,\dot{u}_3)+\lambda(t) \mathcal{F}(t,u_1,u_2)
    \label{augmented_lagrange}
\end{equation}
where we define $u_1:=q$, $u_2:=\eta_A$ and $u_3:=\theta$, \begin{equation}
 \mathcal{F}(t,u_1,u_2):=\int_0^{u_2}
    \sqrt{1+\bigg(u_1{\frac{d \phi}{{d \eta}}}\bigg)^2}\text{d}\eta-L=0
    \label{eq:lengthconst}
\end{equation}
and $\lambda$ is a time-dependent Lagrange multiplier.
\\
\\
Using Eq.\ref{augmented_lagrange}, we arrive at the following compact form of dynamic equations, where $\textbf{q}=[q,\eta_A,\theta]^T$: 
\begin{align}
    \textbf{D}(\textbf{{q}})\ddot{\textbf{q}}+\textbf{C}(\textbf{{q}},\dot{\textbf{q}})\dot{\textbf{q}}+\textbf{R}(\textbf{{q}})\dot{\textbf{q}}+\textbf{G}({\textbf{q}})=\textbf{B}({\textbf{q}})\tau+\lambda\textbf{A}({\textbf{q}})
    \label{eq:dynamics}
\end{align}
or
\begin{equation}
\begin{aligned}
&
\begin{bmatrix}
D_{11} & D_{12} & D_{13} \\
D_{21} & D_{22} & D_{23} \\
D_{31} & D_{32} & D_{33}
\end{bmatrix}
\begin{bmatrix}
\ddot q\\
\ddot{\eta}_A\\
\ddot{\theta}
\end{bmatrix}
\\[0.6em]
&+
\begin{bmatrix}
C_{11}\dot{\eta}_A + C_{14}\dot q &
C_{12}\dot{\theta} + C_{15}\dot{\eta}_A &
C_{13}\dot q + C_{16}\dot{\theta} \\
C_{21}\dot{\eta}_A + C_{24}\dot q &
C_{22}\dot{\theta} + C_{25}\dot{\eta}_A &
C_{23}\dot q + C_{26}\dot{\theta} \\
C_{31}\dot{\eta}_A + C_{34}\dot q &
C_{32}\dot{\theta} + C_{35}\dot{\eta}_A &
C_{33}\dot q + C_{36}\dot{\theta}
\end{bmatrix}
\begin{bmatrix}
\dot q\\
\dot{\eta}_A\\
\dot{\theta}
\end{bmatrix}
\\[0.6em]
&+
\begin{bmatrix}
C_q & 0 & 0\\
0 & 0 & 0\\
0 & 0 & C_\theta
\end{bmatrix}
\begin{bmatrix}
\dot q\\
\dot{\eta}_A\\
\dot{\theta}
\end{bmatrix}
+
\begin{bmatrix}
G_1\\
G_2\\
G_3
\end{bmatrix}
=
\begin{bmatrix}
0\\
0\\
1
\end{bmatrix}\tau
+
\lambda
\begin{bmatrix}
A_1\\
A_2\\
0
\end{bmatrix}
\end{aligned}
\end{equation}
     
    

All the coefficients are given in the Appendix.

\section{Analysis: Multiple Equilibria}

\subsection{Analysis of Equilibria using AMM}

\noindent To obtain equations of equilibrium under the condition of fixed hub angle, we set the higher order derivatives in Eq.\ref{eq:dynamics} to zero. Further elimination of $\lambda$ from the resulting equations yield the following condition:  
 \begin{equation}
        A_1G_2-A_2G_1 = 0
        \label{equilibrium}
    \end{equation}
\noindent According to Euler's buckling theory \cite{LIEW2017141}, the first critical buckling load (corresponding to the first mode of buckling) is given by
\begin{equation}
P_{critical}= \frac{\pi^2EI}{4L^2}
\label{criticalMass}
\end{equation}
For developing the main idea in this paper, we consider the tip mass up to a value such that $Mg$ is more than the first $P_{critical}$ but below the second critical load. This is consistent with our AMM analysis that considers only the first mode. The following cases would now be discussed: 
\subsection{Case 1: $M < M_{critical}$}
 \noindent When the tip mass is less than the critical value, the beam does not buckle, and the trivial configuration ($q=0$) remains a stable equilibrium position. 
  
\subsection{Case 2: $M > M_{critical}$}
\noindent When the tip mass exceeds the first critical value, the system undergoes buckling, and shows very large deflections. The equilibrium states observed for this case depend on both, the buckling load and the hub angle. 
\\
\\
For a given load, with hub angle $\theta = 0$ (i.e. gravity vector aligned with $e_n$), one central unstable equilibrium and two symmetric side equilibria are obtained as solutions of Eq.\ref{equilibrium}. The case corresponds to two symmetric potential energy (PE) wells (in the PE vs q graph as plotted for our case in Fig.\ref{PE_plot}). However, as the hub angle increases, this symmetry is broken, and the solution of Eq.\ref{equilibrium} yields two asymmetric stable equilibria and one unstable equilibrium. Potential energy exhibits local minima at these two side equilibrium positions (two wells) and a local maximum at the unstable equilibrium, different from the trivial one. For this case, the observation of the $A_1G_2-A_2G_1$ graph against q reveals three intersection points with the q-axis (see, for example, Fig.\ref{eq_exp}, plotted specifically for our system parameters). As the hub angle increases further, a critical value (say $\theta_0$) is reached where one of the potential energy wells vanishes, and we get a saddle point behavior. This behavior is characterized,  in addition to Eq.\ref{equilibrium}, by 
\begin{equation}
    d(A_1G_2-A_2G_1)/ dq = 0 \text{ or }
    d(A_1G_2-A_2G_1)/ d\eta_A = 0
    \label{critcal_condition}
\end{equation}
Beyond the critical $\theta_0$, only one stable equilibrium solution is obtained, and the other two solutions vanish. 
Furthermore, as the load increases, the critical $\theta_0$ would increase. 

\subsection{Equilibria using the Elastica Theory}
\noindent Elastica theory by Borgh \cite{borg1990fundamentals} presents a way to carry out accurate static analysis of beams, using variables along the length and hence avoids the use of length constraint. We use elastica analysis to compare static results obtained using AMM. 
Let $\psi(s)$ denote the tangent angle measured from the radial line of the hub,
and let $s \in [0,L]$ be the arc–length coordinate along the beam. We define $P=Mg$ as the known constant force acting on the beam as per the pure Elastica theory. For an inextensible, unshearable, planar beam with bending stiffness $EI$, using the
moment–curvature relation, the static equilibrium of the elastica yields
\begin{equation}
    EI\,\psi''(s) + P \sin \psi(s) = 0
    \label{eq:elastica-ODE}
\end{equation}
which is a nonlinear second–order differential equation. 
\\
Multiplying~\eqref{eq:elastica-ODE} by $\psi'(s)$ and integrating gives the
standard first integral:
\begin{equation}
    \frac{1}{2} EI\,\psi'(s)^{2}
    = P\!\left(\cos\psi(s)-\cos\alpha\right)
    \label{eq:first-integral}
\end{equation}
where $\alpha$ is a constant determined by boundary conditions
(the maximum angular deviation of the elastica).

\noindent By using trigonometric relations as in \cite{borg1990fundamentals}, we can relate variables s, $\psi$, x, and y. The integral becomes an elliptic integral of the first kind.
\begin{equation}
    s = 2\sqrt{\frac{EI}{P}}
    \int_{\psi_0}^{\psi}
    \frac{d\eta}{
      \sqrt{1 - k^{2}\sin^{2}\eta}}
    \label{eq:elliptic-F}
\end{equation}
where $k^{2} = \sin^{2}\!\left(\frac{\alpha}{2}\right)$
The unknown parameter $\alpha$ is determined from the end boundary condition. Once $\psi(s)$ is known, the beam centerline is reconstructed from
\begin{align}
  \eta(s) &= \int_{0}^{s} \cos \psi(\xi)\,d\xi \\
    \zeta(s) &= \int_{0}^{s} \sin \psi(\xi)\,d\xi
    \label{eq:centerline}
\end{align}
followed by a rotation into the global coordinate frame of the hub
\begin{align}
    x(s) &= (R+\eta)\sin\theta + \zeta\cos\theta\\
    y(s) &= (R+\eta)\cos\theta - \zeta\sin\theta
\end{align}
In the following section, we discuss the results of our static analysis and validate them experimentally. We also present a case study of dynamic simulation, which establishes the model's efficacy in capturing nonlinearities. 
\section{Results and Discussion}
\subsection{System Parameters}
\noindent Table \ref{table1} summarizes the key parameters employed in both the numerical simulations and experimental validation. These parameters were chosen to ensure that the load is greater than the first critical buckling load and much less than the second critical buckling load as per our assumptions in the theoretical analysis. 
\begin{table}[h]
\centering
\renewcommand{\arraystretch}{1.6} 
\setlength{\tabcolsep}{20pt} 
\caption{Parameter List}
\label{table1}
\begin{tabular}{|l|l|}
\hline
\textbf{Parameter} & \textbf{Value} \\
\hline
Tip Mass (M) & 0.04 kg\\
Beam Length (L) & 0.315 m\\
Hub Radius (R) & 0.03 m \\
Young's Modulus (E) & 130 GPa\\
Beam Density ($\rho$)& 8400 $kg/m^3$\\
Damping Coefficient of the beam ($C_q$) & 0.000134 $kg/s$\\
\hline
\end{tabular}
\end{table}

\vspace{-12pt}
\subsection{State Dependent Multiple Equilibria}
\begin{figure}
     \centering
     \begin{subfigure}{0.4\textwidth}
         \centering
         \includegraphics[width=1\textwidth]{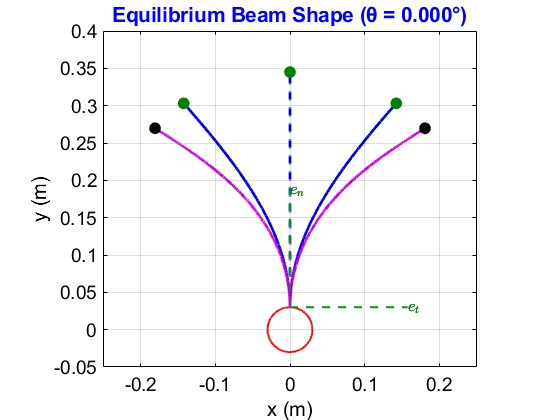}
         \caption{}
         \label{fig2a}
     \end{subfigure}
     \begin{subfigure}{0.4\textwidth}
         \centering
         \includegraphics[width=1\textwidth]{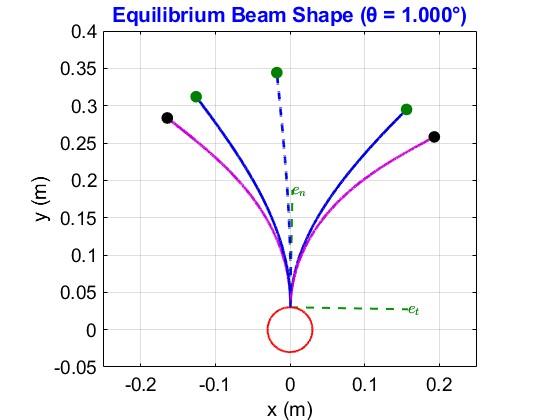}
         \caption{}
         \label{fig2b}
     \end{subfigure}
     \begin{subfigure}{0.4\textwidth}
         \centering
         \includegraphics[width=1\textwidth]{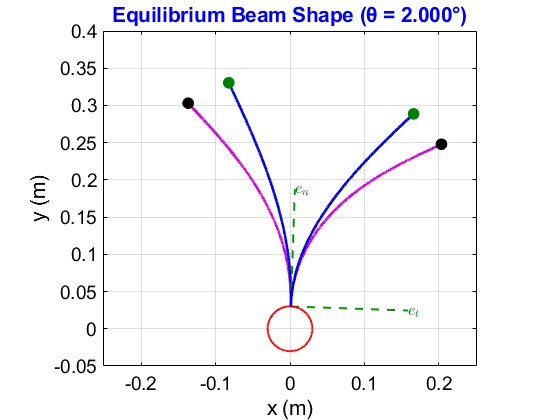}
         \caption{}
         \label{fig2c}
     \end{subfigure}
     \begin{subfigure}{0.4\textwidth}
         \centering
         \includegraphics[width=1\textwidth]{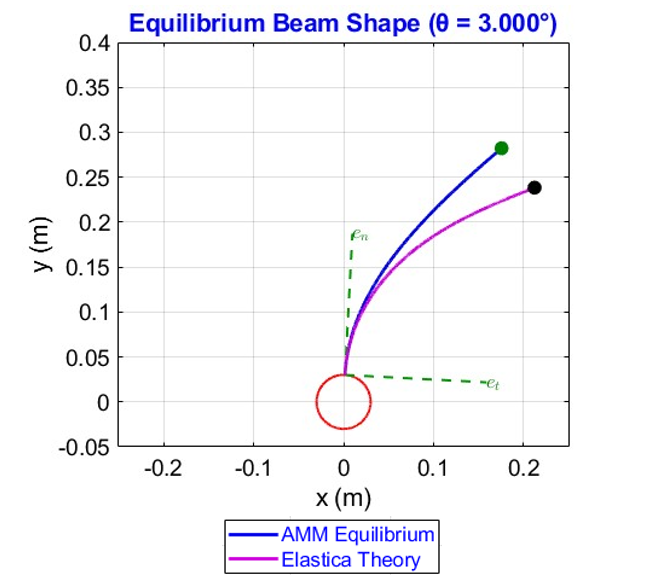}
         \caption{}
         \label{fig2d}
     \end{subfigure}
        \caption{Evolution of state-dependent equilibria of the system}
        \label{elastica}
\end{figure}

\noindent Simulation results in Fig.\ref{elastica}, with parameters defined in Table\ref{table1}, using Eq.\ref{equilibrium} and corresponding other equations, systematically capture the proposed state dependence of the system's equilibria. Initially, when the hub angle is zero (Fig. \ref{fig2a}), we observe symmetric stable equilibria on both sides of the vertical axis. Here, the $e_n$ axis is perfectly aligned with the vertical gravity vector. We also observe that the AMM solution depicts a stiffer beam than the elastica solution, because of the assumption of only one mode in the analysis. The corresponding plot of Eq.\ref{equilibrium} shows these solutions as symmetric intersections on the q-axis as seen in Fig.\ref{eq_exp}, with one central point at $q = 0$. 
As the hub angle is increased further (Fig.\ref{fig2b},\ref{fig2c}, Fig.\ref{eq_exp}), the symmetry of equilibria is broken, but nature still remains the same: two side stable equilibria and one unstable. For further increase beyond the critical hub angle (Fig.\ref{fig2d}, Fig.\ref{eq_exp}), only one stable equilibrium persists, and other equilibria vanish. 

\begin{figure}[h!]
    \centering
    \includegraphics[width=0.5\textwidth]{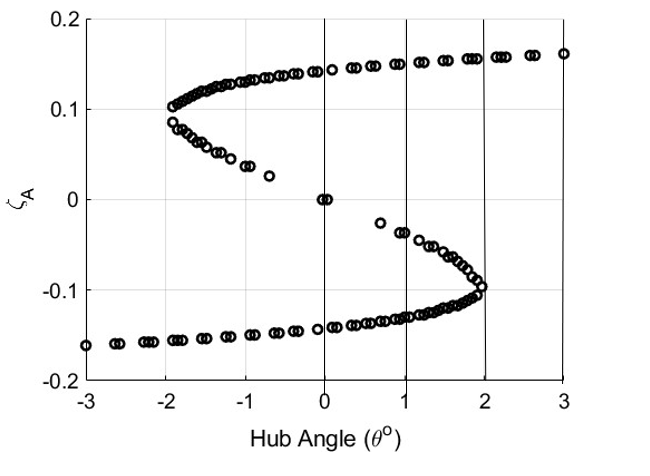}
    \caption{Existence of 3 equilibria at $0\degree,1\degree$, 2 equilibria at $\approx2\degree$ and 1 equilibrium at $3\degree$}
    \label{s_curve}
\end{figure}

\begin{figure}[h!]
    \centering
    \includegraphics[width=0.5\textwidth]{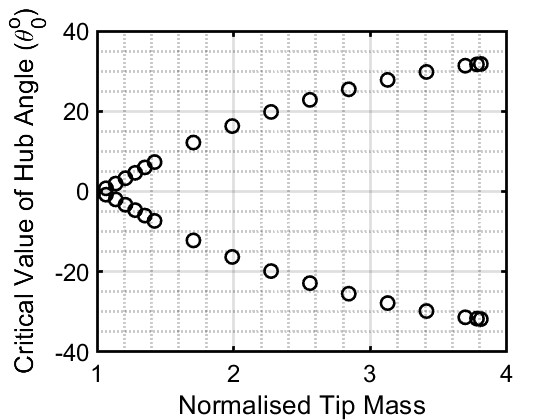}
    \caption{{ $\theta_0\degree$ vs $M$/$M_{critical}$}}
    \label{Mass_variation}
\end{figure}

\begin{figure}
\centering
 \begin{subfigure}{0.5\textwidth}
    \centering
    \includegraphics[width=1\textwidth]{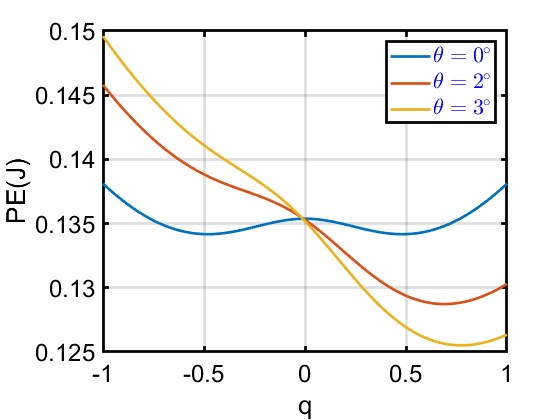}
    \caption{}
    \label{PE_plot}
     \end{subfigure}
\begin{subfigure}{0.5\textwidth}
         \centering
         \includegraphics[width=1\textwidth]{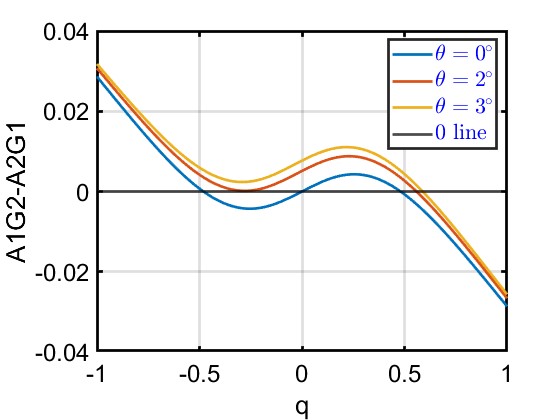}
         \caption{}
         \label{eq_exp}
     \end{subfigure}
     \caption{(a) value of $A_1G_2-A_2G_1$ from equilibrium condition (Eq.\ref{equilibrium}) (b) Potential energy (PE) of the system 
}
\end{figure}

\noindent Simulation results in Fig.\ref{s_curve} show equilibrium displacement of the tip point $\zeta_A$ as hub angle is varied. The top and bottom branches of the "s" curve correspond to stable equilibria, while the middle part corresponds to unstable equilibria. Two saddle nodes observed around $\theta_0 = \pm2$ are solutions of Eq.\ref{critcal_condition}. 
\\
\\
\noindent As the tip mass is increased beyond the first critical buckling mass, the critical hub angle beyond which the two equilibria vanish varies as shown in Fig.\ref{Mass_variation}. This behaviour indicates a nonlinear sensitivity of the equilibrium state to changes in tip mass under post-buckling conditions. We know from \cite{10.1115/1.4026831}, that we observe more equilibria as we consider more modes in our derivation of the dynamics using AMM and go beyond critical loads of higher orders. We expect a similar kind of behaviour from our system too, for which the appropriate critical angles can be calculated when considering higher modes.

\subsection{Dynamic Simulation}

\begin{figure}[h!]
     \centering
     \begin{subfigure}{0.5\textwidth}
         \centering
         \includegraphics[width=1\textwidth]{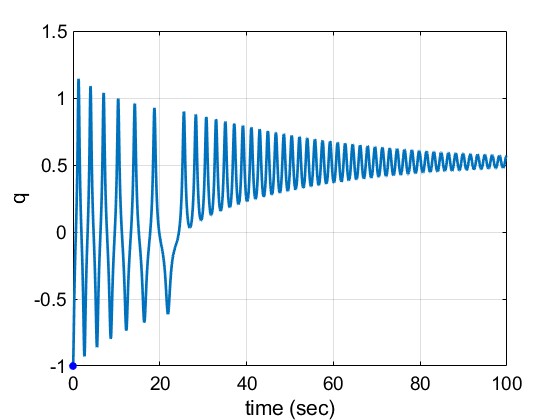}
         \caption{}
         \label{}
     \end{subfigure}
     \begin{subfigure}{0.5\textwidth}
         \centering
         \includegraphics[width=1\textwidth]{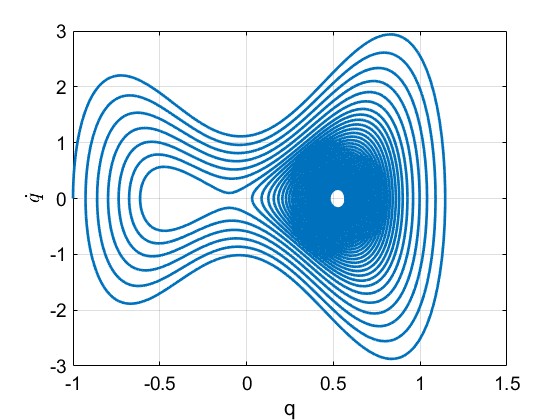}
         \caption{}
         \label{phase_plot}
     \end{subfigure}
        \caption{Dynamic Simulation for fixed hub angle $\theta = 1\degree$ }
        \label{dynamics}
\end{figure}

\noindent One interesting simulation result case is presented in this section to get an indicative glimpse of the highly nonlinear behavior of the system captured well by the proposed model based on AMM. We fix the hub at $1\degree$ and give initial excitation to the tip mass with q = -1. A phase plot of q in Fig.\ref{phase_plot} shows initial movement of mass over all stable and unstable equilibria and settling into one stable equilibrium eventually. 

\subsection{Experimental Results}
\begin{figure}[h!]
\centering
\includegraphics[width=0.44\textwidth]{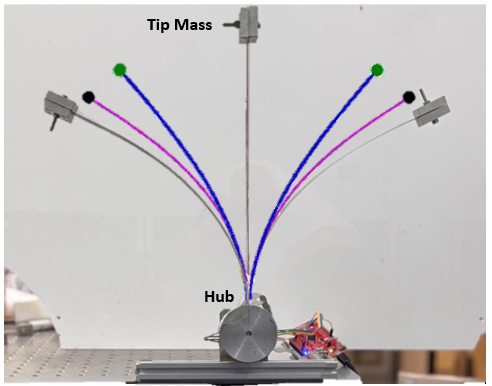}
    \caption{{Comparison of Multiple equilibria at $\theta = 0\degree$} from experiments and simulations}
    \label{0_upright}
\end{figure}

\noindent An experimental set-up was developed to compare and validate theoretical findings. The hub is manufactured from aluminum alloy, and the beam from BeCu alloy (high fatigue resistance) using precision EDM cutting. Arrangements were made to attach the beam such that the beam protrudes out radially from the hub centre. An encoder- AMT102 D0500-I8000s was used to measure the angular displacement of hub.
A customised clamp is manufactured to lock the hub in the required position. It was ensured that the $0\degree$ on the hub corresponds to the vertically upward position. Static stable equilibrium positions were captured for the following hub angles:
 $-3\degree$, $-2.54\degree$, $-2.36\degree$, $-2.18\degree$, $-2\degree$, $-1\degree$, $0\degree$, $1\degree$, $2\degree$, $2.18\degree$, $2.36\degree$, $2.54\degree$, and $3\degree$
\\
\\
Fig. \ref{0_upright} shows the system at $0\degree$ with the beam in all its equilibria. The experimental images have been superposed to depict the existence of multiple equilibria. For the purpose of comparison, we have also superimposed the simulation results for equilibrium shapes derived from both, the AMM and Elastica method, at $0\degree$. We can see clearly that the system is much stiffer in simulation.
\noindent The software Tracker Version 6.3.1 was used for processing all images to finally extract the distances from the rotated local coordinate system ($\eta$ and $\zeta$) for ready comparison with corresponding simulation results.

\subsection{Comparison of experimental results with simulation results}

\begin{figure}[h!]
\centering
    \includegraphics[width=0.44\textwidth]{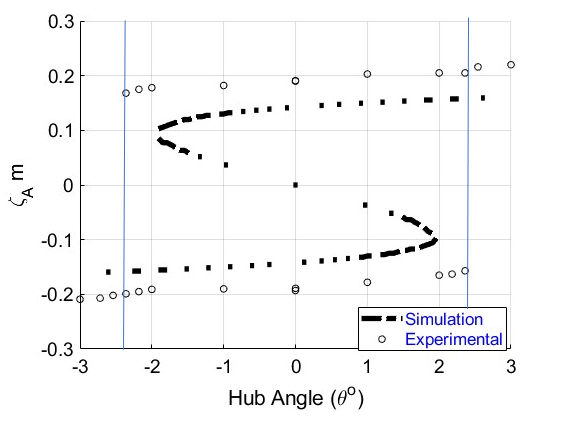}
    \caption{{Comparison between Simulation and Experiments}}
    \label{Comp}
\end{figure}
\noindent Fig.\ref{Comp} shows comparison of simulated vs experimental tip mass equilibrium displacement $\zeta_A$ as a function of hub angle. This plot is obtained by repeating experiments using the setup in Fig.\ref{0_upright} for the hub angles mentioned above and processing images taken in the steady state. The unstable equilibrium states (the middle part of the "s" curve) are infeasible to capture. 
\\
\\
We see that the simulation and experimental data points are close to each other. It can be observed that, for experiments conducted on the positive side, i.e., when the beam bends in the direction of hub rotation, the value of $\eta$ obtained experimentally is lower than that predicted by simulation, whereas the value of $\zeta$ is higher. We observe similar results of magnitude of $\eta$ and $\zeta$ for the beam bent in the opposite direction. 
\\
\\
These results corroborate our earlier observation that the AMM model, owing to the first mode assumption, is stiffer as compared to the experimental system with the same parameters. Hence critical hub angle in the experimental case is also higher. 
\begin{equation}
    \theta_{0,sim} = 2.03\degree ; \theta_{0,exp} \approx 2.36\degree
\end{equation}
\noindent A case of dynamic experimentation, with the hub angle fixed at $1\degree$ can be observed using the following {\href{https://drive.google.com/file/d/1pXB6M0tWylT_OWRLjCdONx5znw5gGwPj/view}{\color{blue}link}. From this experiment, we can see that the pendulum switches between the two equilibria and even slows down near the unstable equilibria. Finally, due to damping,
it slows down and chooses one from the two stable equilibria and oscillates before settling.
\section{Conclusion}
\noindent This work considers the dynamic modeling of a flexible inverted pendulum with a tip mass on a rotary hub system, focusing particularly on continuously state-dependent and eventually vanishing multiple equilibria of this system. This interesting behavior is attributed to an additional change in potential energy of the system as the hub angle changes on either side of the zero. A dynamic model of the system is obtained using the constrained Lagrange formulation, considering large deformations and the assumed modes method framework with only the first mode. Equilibria obtained from the dynamic model, and those from elastica theory are analysed to demonstrate the bifurcation as the hub angle is increased on either side. Moreover, the critical hub angle where the bifurcation occurs is characterized with respect to the normalized tip mass. Theoretical insights are further supported by meticulously carried out experimental results.
\\
\\
Compliant systems with complex nonlinear dynamics with an infinite degree of freedom are challenging yet advantageous to work with in various applications, ranging from soft robotics to MEMS, to surgical robotics, and so on. The insights, reported here for the first time to the best of the authors' knowledge, provide a strong foundation in developing and controlling post-buckled ultra-flexible links and mechanisms in various orientations in these applications.

\section{Interests}
\noindent This work is purely inspired by the previous work of P.Gandhi on flexible systems. Part of the funding from a research project is used for exploring experimental aspects.

\section{Data Availability}
\noindent No datasets were generated or analyzed during the current study.
\section{Appendix}

\begin{multline}
    D_{11}=\rho\int^{\eta_A}_0 \phi^2\sqrt{1+q^2\phi'^2}d\eta +M\phi(\eta_A)^2
\end{multline}
\begin{equation}
    D_{12}=Mq\phi(\eta_A)\phi'(\eta_A)
\end{equation}
\begin{multline}
    D_{13}=\rho\int^{\eta_A}_0 \phi(R+\eta)\sqrt{1+q^2\phi'^2}\text{d}\eta +M(R+\eta_A)\phi(\eta_A)
\end{multline}
\begin{equation}
    D_{21}=Mq\phi(\eta_A)\phi'(\eta_A)
\end{equation}
\begin{equation}
    D_{22}=M(1+q^2\phi'(\eta_A)^2)
\end{equation}
\begin{equation}
    D_{23}=Mq((R+\eta_A)\phi'(\eta_A)-\phi(\eta_A))
\end{equation}
\begin{multline}
    D_{31}={\rho}\int^{\eta_A}_0
    \sqrt{1+q^2\phi'^2}{[}(R+\eta)\phi{]}\text{d}\eta\\+M(R+\eta_A)\phi(\eta_A)
\end{multline}
\begin{equation}
    D_{32}=Mq(-\phi(\eta_A)+(R+\eta_A)\phi'(\eta_A))
\end{equation}
\begin{multline}
    D_{33}={\rho}\int^{\eta_A}_0
    \sqrt{1+q^2\phi'^2}{[}q^2\phi^2+(R+\eta)^2{]}\text{d}\eta+\mathcal{J}
    \\+Mq^2\phi(\eta_A)^2+M(R+\eta_A)^2
\end{multline}
\begin{multline}
    C_{11}=2M\phi(\eta_A)\phi'(\eta_A)+\rho\phi(\eta_A)^2\sqrt{1+q^2\phi'(\eta_A)^2}
\end{multline}
\begin{multline}
    C_{12}=2M\phi(\eta_A)+\rho\phi(\eta_A)(R+\eta_A)\sqrt{1+q^2\phi'(\eta_A)^2}
\end{multline}
\begin{equation}
    C_{13}=0
\end{equation}
\begin{equation}
     C_{14}=\frac{\rho}{2}\int^{\eta_A}_0\frac{q\phi(\eta)^2\phi'(\eta)^2}{\sqrt{1+q^2\phi'(\eta)^2}}\text{d}\eta
\end{equation}
\begin{equation}
    C_{15}=Mq\phi(\eta_A)\phi''(\eta_A)
\end{equation}
\begin{multline}
    C_{16}=-Mq\phi(\eta_A)^2\\-\frac{q\rho}{2}\int^{\eta_A}_0\frac{\big{(}q^2\phi^2+(R+\eta)^2\big{)}\phi'^2+2\phi^2\big{(} 1+q^2\phi'^2\big{)} }{\sqrt{1+q^2\phi'^2}}
    \text{d}\eta
\end{multline}
\begin{equation}
    C_{21}=2Mq\phi'(\eta_A)^2
\end{equation}
\begin{equation}
    C_{22}=0
\end{equation}
\begin{multline}
    C_{23}=-2M\phi(\eta_A)\\-\rho(R+\eta_A)\phi(\eta_A)\sqrt{1+q^2\phi'(\eta_A)^2}
\end{multline}
\begin{equation}
    C_{24}=-\frac{1}{2}\rho\sqrt{1+q^2\phi'(\eta_A)^2}\phi(\eta_A)^2
\end{equation}
\begin{equation}
    C_{25}=Mq^2\phi'(\eta_A)\phi''(\eta_A)
\end{equation}
\begin{multline}
    C_{26}=-M(R+\eta_A)-Mq^2\phi(\eta_A)\phi'(\eta_A)\\-\frac{1}{2}\rho \sqrt{1+q^2\phi'(\eta_A)^2}{[}\phi(\eta_A)^2q^2+(R+\eta_A)^2{]}
\end{multline}
\begin{multline}
    C_{31}=(R+\eta_A)\Big[\rho\sqrt{1+q^2\phi'(\eta_A)^2}\phi(\eta_A)+2M\phi'(\eta_A)\Big]
\end{multline}
\begin{multline}
    C_{32}=\rho\sqrt{1+q^2\phi'(\eta_A)^2}(q^2\phi(\eta_A)^2+(R+\eta_A)^2)+\\2Mq^2\phi(\eta_A)\phi'(\eta_A)+2M(R+\eta_A)
\end{multline}
\begin{multline}
    C_{33}=2Mq\phi(\eta_A)^2 \\+ {\rho}\int^{\eta_A}_0\frac{q\phi'^2(q^2\phi^2+(R+\eta)^2)}{\sqrt{1+q^2\phi'^2}}+2q\phi^2\sqrt{1+q^2\phi'^2}\text{d}\eta
\end{multline}
\begin{equation}
    C_{34}=\rho\int^{\eta_A}_0\frac{q\phi(\eta)\phi'(\eta)^2(R+\eta)}{\sqrt{1+q^2\phi'(\eta)^2}}\text{d}\eta
\end{equation}
\begin{equation}
    C_{35}=Mq(R+\eta_A)\phi''(\eta_A)
\end{equation}
\begin{equation}
    C_{36}=0
\end{equation}
\begin{multline}
    G_1=\frac{EI}{2}\int^{\eta_A}_0\frac{ q(2-3q^2\phi'^2)\phi''^2}{({1+q^2\phi'^2})^{7/2}}\text{d}\eta- Mg\phi(\eta_A)\sin{(\theta)}
\end{multline}

\begin{multline}
    G_2=Mg\Big[\cos(\theta)-q\phi'(\eta_A)\sin{\theta}\Big]+\frac{EIq^2\phi''(\eta_A)^2}{2(1+q^2\phi'(\eta_A)^2)^{5 /2}}
\end{multline}

\begin{multline}
    G_3=-Mg(q\phi(\eta_A)\cos{(\theta)}+(R+\eta_A)\sin{(\theta)})
\end{multline}
\begin{equation}
    A_1=\int^{\eta_A}_0 \frac{q\phi'(\eta)^2}{\sqrt{1+q^2\phi'(\eta)^2}}\text{d}\eta
\end{equation}
\begin{equation}
    A_2=\sqrt{1+q^2\phi'(\eta_A)^2}
\end{equation}


\bibliographystyle{IEEEtran}
\bibliography{ref}

\end{document}